\documentclass[fleqn,twoside]{article}
\usepackage[headings]{espcrc2}
\usepackage{moreverb,axodraw}
\newcommand\greyed[1]{\special{ps: .7 setgray}#1\special{ps: 0 setgray}}
\renewcommand\listingoffset{.5em}
\renewcommand\listinglabel[1]{%
  \llap{\scriptsize\sffamily\greyed{\the#1}\hskip\listingoffset}\relax}

\newcommand{\eg}{e.g.\ }
\newcommand{\ie}{i.e.\ }
\newcommand\Code[1]{\texttt{#1}}

\begin{document}

\author{T. Hahn\address{%
	Max-Planck-Institut f\"ur Physik
\hfill MPP-2011-88 \\
	F\"ohringer Ring 6, D--80805 Munich, Germany
\hfill PACS: 02.70.Wz, 07.05.Bx, 07.05.Wr}}

\title{The High-Energy Physicist's Guide to MathLink}

\begin{abstract}
MathLink is Wolfram Research's protocol for communicating with the
Mathematica Kernel and is used extensively in their own Notebook 
Frontends.  The Mathematica Book insinuates that linking C programs 
with MathLink is straightforward but in practice there are quite a 
number of stumbling blocks, in particular in cross-language and 
cross-platform usage.  This write-up tries to clarify the main issues 
and hopefully makes it easier for software authors to set up Mathematica
interfacing in a portable way.
\end{abstract}

\maketitle

\section{Introduction}

Wolfram Research's Mathematica is conceptually set up in two pieces, the 
Kernel (the computational engine) and the Notebook Frontend (the GUI, 
also used for formatting, rendering, etc.).  The MathLink protocol is 
the means of communicating with a running Mathematica Kernel, both from 
the Kernel invoking an external program and an external program invoking 
the Kernel, where most commonly the `external program' is the Frontend.  
If necessary, the communication can take place across the internet
through a TCP connection.

The MathLink SDK is installed together with Mathematica and its API is 
documented in Mathematica's Documentation Center (`Help' menu), which 
describes the operations for linking a C program with MathLink as rather 
trivial.  In everyday work this is true only for simple C programs, 
however.  Cross-language or cross-platform use in particular is fairly 
non-straightforward.

This article does not cover interpreted languages such as Java or Python 
for some of which own interfaces exist (\eg J/Link).

MathLink has been a moving target over the Mathematica versions and 
platforms.  The scripting code given herein represents the state of 
affairs up to Mathematica 8, MacOS 10.6, Windows 7/Cygwin 1.7, and, of 
course, any flavour of Linux.  The scripts are described in the 
following for better understanding but can largely be used as black 
boxes.  All code is available for download at 
\Code{http://feynarts.de/mathlink/}.  The line numbers of the code 
excerpts in this text do not necessarily correspond to the actual
scripts.

\section{General setup}

There are various ways of setting up a MathLink program, some of which 
depend on the programming environment (\eg XCode).  We concentrate here 
on the `standard' MathLink template file (\Code{.tm}) which is portable 
across all platforms.

The following is only a brief introduction to the template file format.  
For a more thorough treatment and reference please see the Mathematica 
Help Browser under `.tm file'.

A template file consists of three parts:

\noindent
1. A header identifying the functions visible from Mathematica,
for example
\begin{listing}{1}
:Begin:
:Function: a0
:Pattern: A0[m_, opt___Rule]
:Arguments: {N[m],
   N[Delta /. {opt} /. Options[A0]],
   N[Mudim /. {opt} /. Options[A0]]}
:ArgumentTypes: {Real, Real, Real}
:ReturnType: Real
:End:

:Evaluate: Options[A0] =
   {Delta -> 0, Mudim -> 1}
\end{listing}
Ostensibly, argument and option processing happens at this point, and 
the C function receives exactly the quantities given under 
\Code{:Arguments:}.

\medskip

\noindent
2. C code implementing those functions,
\begin{listingcont}
#include "mathlink.h"

static double a0(const double m,
const double delta, const double mudim) {
  return (m == 0) ? 0 :
    m*(1 - log(m/mudim) + delta);
}
\end{listingcont}

\medskip

\noindent
3. A main function which might set up global variables, invoke 
initialization routines, etc., but eventually hands control over to
\Code{MLMain}:
\begin{listingcont}
int main(int argc, char **argv) {
  return MLMain(argc, argv);
}
\end{listingcont}
\Code{MLMain} returns if the MathLink program is uninstalled either by
an explicit \Code{Uninstall} or by quitting the corresponding 
Mathematica Kernel.  After that \Code{main} may want go through some
finalization procedure (\eg closing files) before finally terminating.

The MathLink API functions are documented in Mathematica's Help Browser,
including a tutorial under
`tutorial/MathLinkAndExternal\-Pro\-gram\-CommunicationOverview'.

Compiling such a program should in principle be as easy as substituting 
\Code{mcc} for \Code{cc} on the command line, as in:
\begin{verbatim}
mcc -o mlprog mlprog.tm
\end{verbatim}
More on compilation in Sect.\ \ref{sect:mcc} below, however.

Once the MathLink program has been built successfully, it can be 
installed in Mathematica using
\begin{verbatim}
Install["mlprog"]
\end{verbatim}
Unless invoked with an explicit path, the program has to be on either 
the system \Code{PATH} or on Mathematica's \Code{\$Path}.

Alternately, start the MathLink program on the command line:
\begin{verbatim}
> ./mlprog
Create link:
\end{verbatim}
Choose an arbitrary string as a name for the link and enter it here.  
In Mathematica, type
\begin{verbatim}
Install[LinkConnect["linkname"]]
\end{verbatim}
where \Code{linkname} is the chosen name, to establish the connection.  
In this way, \Code{mlprog} can also be started in the debugger, which
is the routine way of debugging MathLink programs.

If the link target is on another machine, start the MathLink program
with
\begin{verbatim}
./mlprog -linkname port \
         -linkprotocol TCPIP
\end{verbatim}
where \Code{port} is an integer larger than 1024 (a port number not in 
the reserved range) and connect from Mathematica via
\begin{verbatim}
Install[LinkConnect["port@host",
  LinkProtocol -> "TCPIP"]]
\end{verbatim}

\section{Linking}
\label{sect:linking}

Unix (Linux, MacOS, \dots) linkers are one-pass linkers (except when 
linking shared libraries).  What this means is that if library 
\Code{libneed.a} references symbol X and \Code{libprovide.a} provides X, 
the following command nevertheless fails to resolve X:
\begin{verbatim}
cc ... -lprovide -lneed
\end{verbatim}
because the linker does not `go back' to earlier libraries.  If 
necessary, \Code{-lprovide} has to appear several times on the command 
line.

When providing ready-made executables to the general public, I strongly 
advertise statically linking the executable as far as possible 
(\Code{-static} in gcc, \Code{-st} in mcc).  A statically linked 32-bit 
x86 Linux executable, for example, should run on essentially all Linux 
flavours currently available and will likely do so 10 years from now.

Besides, executables dynamically linked with the MathLink libraries 
require the MathLink library directory to be included on the library 
path (\Code{LD\_LIBRARY\_PATH} or \Code{/etc/ld.so.conf}) when invoked 
outside of Mathematica or in Mathematica versions before 6.

On MacOS and Windows it is not possible to statically link the system 
libraries, as \eg Apple reserves the right to change their system 
library in an upward-compatible way.  One should try to statically link 
at least the external libraries of the compiler(s) used, such as Fortran 
run-time libraries, so as to make the MathLink executable independent of 
those compilers.  In the case of gcc, replace \Code{-static} by 
\Code{-static-libgcc}.

\section{The \Code{mcc} Compiler}
\label{sect:mcc}

The MathLink C compiler, \Code{mcc}, is a shell script which supposedly 
replaces \Code{cc}, the C compiler, in MathLink applications using 
approximately the same command line.  Even after many e-mails to Wolfram 
Support, \Code{mcc} still fails to take care of library ordering, 
however.

Luckily, \Code{mcc} observes the \Code{CC} and \Code{CXX} environment 
variables, making it possible to substitute the actual C/C++ compilers 
with a shell script that corrects the library ordering.  Such a script 
would look like
\begin{listing}{1}
#! /bin/sh
# script to compile C programs that are
# linked against Fortran libraries

args=
objs=
ldflags=
fldflags=
compileonly=

cc="${REALCC:-cc}"
cxx="${REALCXX:-c++}"
test `basename $0` = f++ && cc="$cxx"

while test $# -gt 0 ; do
  case "$1" in
  -st | -b32 | -b64)
        ;; # ignore mcc-specific flags
  -arch)
        shift ;;
  -lstdc++)
        cc="$cxx" ;;
  -[Ll]* | -Wl*)
        ldflags="$ldflags '$1'" ;;
  *.tm.o)
        objs="'$1' $objs" ;;
  *.a | *.o | *.so)
        objs="$objs '$1'" ;;
  *.cc)
        args="$args '$1'"
        cc="$cxx" ;;
  -c)
        compileonly="-c" ;;
  -o)
        args="$args -o '$2'"
        shift ;;
  *)
        args="$args '$1'" ;;
  esac
  shift
done

eval "set -x ; exec $cc $args \
${compileonly:-$objs $ldflags $fldflags}"
\end{listing}
Lines 5--12 initialize a few variables.  Since the environment variable 
\Code{CC} is obviously taken, we need \Code{REALCC} and \Code{REALCXX} 
as replacements for choosing the actual C and C++ compilers, 
respectively.

This script is named \Code{fcc} (`C compiler for linking with Fortran') 
and is symlinked to \Code{f++}.  When invoked as \Code{f++}, the C++ 
compiler is taken as default (line 13).

The \Code{while} loop starting in line 15 runs over all arguments and 
categorizes them: object files into \Code{objs}, libraries and linker 
flags into \Code{ldflags}, a \Code{-c} into \Code{compileonly}, all 
other arguments (\eg source files, compiler flags) into \Code{args}.

An \Code{-arch x} argument is removed (line 17--18) because \Code{mcc} 
tries to build the executable for all admissible platforms, \eg on Mac 
for both Intel and PPC, and this conflicts during linking if an external 
library does not contain object code for all the given architectures.  
An explicit \Code{-lstdc++} has to be removed likewise (line 21--22) as 
it would render a possible \Code{-static-libstdc++} flag useless.  The 
explicit (re-)quoting and the final \Code{eval} (line 43) are necessary 
to treat arguments with spaces properly.

Finally, the actual compiler is invoked with the arguments in the proper 
order (lines 43--44).  If \Code{compileonly} is set, the object files 
and linker flags are omitted.  The \Code{set -x} makes the shell echo 
the actual compiler command line, which is a very good diagnostic if 
anything goes wrong.

The seemingly unused variable \Code{fldflags} can be modified to contain 
external libraries needed for linking with the object files of a 
particular compiler, \eg \Code{-lgfortran} in the case of gfortran (see 
App.\ \ref{app:fldflags}).  In this way, \Code{fcc} serves also as a 
compile script for non-MathLink programs linked with code from another 
compiler.

\bigskip

The invocation of \Code{mcc} must now include the definition of 
\Code{CC} and \Code{CXX}.  Though \Code{fcc} acts largely transparently, 
it is probably not a good idea to set \Code{CC} and \Code{CXX} 
permanently as they are understood by several tools.  In a makefile,
the following syntax might be used:
\begin{verbatim}
FCC = path/to/fcc
mlprog: mlprog.tm
   REALCC="$(CC)" REALCXX="$(CXX)" \
   CC="$(FCC)" CXX="$(FCC)" \
   mcc -o mlprog $(CFLAGS) mlprog.tm \
     $(LDFLAGS) $(LIBS)
\end{verbatim}

\section{Finding \Code{mcc}}
\label{sect:mccpath}

Except on Linux, \Code{mcc} is not installed in a location which is on 
the \Code{PATH}, so we have to find it first.  To this end we make a 
script of the same name, \Code{mcc}, and place that in a directory which 
we append to the \Code{PATH}.  If a true \Code{mcc} is on the 
\Code{PATH}, as in Linux, this gets called.  If not, the substitute 
script gets called.

The substitute script first dispatches to the appropriate OS-specific 
routine:
\begin{listing}{1}
case `uname -s` in
Darwin) macmcc "$@" ;;
CYG*) cygmcc "$@" ;; 
*) defaultmcc "$@" ;;
esac
\end{listing}
Since a naive \Code{find} over the entire hard disk is not feasible,
we resort to heuristics: we search for the Mathematica Kernel in a list 
of typical locations and run that to determine the \Code{\$TopDirectory}, 
which is where the copy of Mathematica to which that Kernel belongs is 
installed.  This is done by the \Code{sdkpath} function (described below)
which receives the name of the Kernel (argument 1) and the list of
possible locations (arguments 2--end).

On MacOS that is all, \ie once the Mathematica directory is known, we 
start \Code{mcc} from that location and exit:
\begin{listingcont}
macmcc() {
  sdkpath MathKernel \
    {/Applications,$HOME/Desktop}/\
Mathematica*/Contents/MacOS
  exec "$sdk/mcc" "$@"
}
\end{listingcont}
The \Code{cygmcc} function for Windows has to do quite a bit more and 
will be discussed in Section~\ref{sect:cygwin} on Cygwin below.  The 
\Code{defaultmcc} function acts as a catch-all.  It tries a few standard
places in case \eg some ignorant system administrator installed 
Mathematica in a location not on the path:
\begin{listingcont}
defaultmcc() {
  sdkpath math \
    /usr/local/bin \
    /usr/local/Wolfram/bin \
    /usr/local/Wolfram/\
Mathematica/*/Executables \
    /opt/Wolfram/bin \
    /opt/Wolfram/\
Mathematica/*/Executables
  exec "$sdk/mcc" "$@"
}
\end{listingcont}

But now the \Code{sdkpath} function:
\begin{listingcont}
sdkpath() {
  mathcmd="$1"
  shift
  mathcmd=`IFS=:
    PATH="$PATH:$*" which $mathcmd`

  eval `"$mathcmd" -run '
Print["sysid=\"", $SystemID, "\""];
Print["topdir=\"", $TopDirectory, "\""];
Exit[]
  ' < /dev/null | tr '\r' ' ' | tail -2`

     # check whether Cygwin's dlltool
     # can handle 64-bit DLLs
  test "$sysid" = Windows-x86-64 && {
    ${DLLTOOL:-dlltool} --help | \
    grep x86-64 > /dev/null || \
      sysid=Windows
  }

  topdir=`cd "$topdir" ; echo $PWD`

  for sdk in \
    "$topdir/SystemFiles/Links/MathLink/\
DeveloperKit/$sysid/CompilerAdditions" \
    "$topdir/SystemFiles/Links/MathLink/\
DeveloperKit/CompilerAdditions" \
    "$topdir/AddOns/MathLink/\
DeveloperKit/$sysid/CompilerAdditions"
  do
    test -d "$sdk" && return
  done

  echo "MathLink SDK not found" 1>&2
  exit 1
}
\end{listingcont}
Lines 24--27: Search for the specified Mathematica Kernel in the given 
list of locations by temporarily adding them to the \Code{PATH}.
Then run the resulting Kernel (line 29) and have it print out (lines 
30--31) the \Code{\$SystemID}, a string identifying the platform, and 
\Code{\$TopDirectory}, the installation directory of that Mathematica 
copy.

On Windows-64, downgrade the \Code{\$SystemID} to 32-bit (line 37--41) 
if Cygwin's \Code{dlltool} professes not to handle 64-bit libraries 
(see Sect.~\ref{sect:cygwin} below).  Canonicalize 
\Code{\$TopDirectory} (line 43), mainly to get rid of Windows-style 
path names (\verb=C:\x\y=).

Go through the list of typical places underneath \Code{\$TopDirectory} 
(lines 46--51) and return the first match (line 53).  If there is no 
match, exit with an error (lines 56--57).

\section{Cygwin}
\label{sect:cygwin}

Cygwin is a Unix-like environment running natively (\ie not virtualized) 
on Windows.  In the sense in which the colloquial `Linux' is more
correctly GNU/Linux (GNU utilities, Linux kernel), Cygwin might be termed 
GNU/Windows.  It is likely the least painful way of porting programs 
from Unix to Windows.

The installation is straightforward even for non-expert users 
(\Code{http://cygwin.com}), though one should review the package 
selection as many tools obviously necessary to build MathLink programs, 
such as \Code{gcc}, \Code{g++}, \Code{make}, are not included in 
the default setup.

The \Code{cygmcc} function starts in much the same way as the 
\Code{macmcc} function, by finding the path to Mathematica:
\begin{listingcont}
cygmcc() {
  sdkpath math \
    "`cygpath '$ProgramW6432'`/\
Wolfram Research/Mathematica"/* \
    "`cygpath '$PROGRAMFILES'`/\
Wolfram Research/Mathematica"/*
\end{listingcont}
\Code{\$ProgramW6432} and \Code{\$PROGRAMFILES} point to the 64- and 
32-bit Applications directory on Windows (``\verb=C:\Program Files=''), 
respectively.  \Code{cygpath} turns Windows-style into Unix-style path 
names (\Code{/cygdrive/c/Program Files}).

From Mathematica 7 on, the MathLink SDK does include Cygwin libraries 
and tools, but these are broken so badly (\eg filename quoting) that not 
even simple programs can be built successfully.  For this reason we skip 
the \Code{cygwin} directory underneath \Code{\$sdk} and move to one of 
the \Code{mldev} directories (\eg \Code{mldev32}):
\begin{listingcont}
for sdk in "$sdk"/m* ; do
  break
done
\end{listingcont}
The chosen \Code{mldev} directory contains native Windows DLLs.  To 
make Cygwin's \Code{ld} accept these, one needs to create a so-called 
`library stub' which contains the library's location and symbol table.  
Cygwin's \Code{dlltool} provides this information and can be used \eg as 
follows:
\begin{listingcont}
cache=MLcyg-cache
test -d $cache || mkdir $cache

MLversion=3
for OSbits in 32 64 ; do
  dllname=ml${OSbits}i$MLversion
  libname="$sdk/lib/${dllname}m.lib"
  test -f "$libname" && break
done

lib="$cache/${dllname}m"
test -f "$lib.a" || {
  ( echo "EXPORTS"
    ${NM:-nm} -C --defined-only \
      "$libname" | \
      awk '/ T [^.]/ { print $3 }'
  ) > "$lib.def"
  ${DLLTOOL:-dlltool} -k \
    --dllname "$dllname.dll" \
    --def "$lib.def" \
    --output-lib "$lib.a"
}
\end{listingcont}
Creating a library stub is a one-time process, so we can store it in a 
cache directory, here named \Code{MLcyg-cache} (lines 68--69).  Lines 
72--76 check which MathLink library (32- or 64-bit) exists in the given 
path.  If the corresponding library stub is not yet in the cache (line 
79) the relevant information is extracted using \Code{nm} and 
\Code{dlltool} and the library stub is created (lines 80--88).

Finally, we have to emulate \Code{mcc}:
\begin{listingcont}
tmp=
args="-DWIN$OSbits -I'$sdk/include'"
for arg in "$@" ; do
  case "$arg" in
  *.tm)
      cp "$arg" "$arg.tm"
      "$sdk"/bin/mprep -lines \
        -o "$arg.c" "$arg.tm"
      tmp="$tmp '$arg.c' '$arg.tm'"
      args="$args '$arg.c'" ;;
  *)
      args="$args '$arg'" ;;
  esac
done

trap "rm -f $tmp" 0 1 2 3 15
eval "set -x ; \
  ${CC:-gcc} $args $lib.a -mwindows"
\end{listingcont}
All \Code{.tm} files are converted to C code using the \Code{mprep} 
utility (lines 96--97).  Note the explicit \Code{cp} in line 95
which is necessary in case the original \Code{.tm} file is a symlink
(symlinks are understood by Cygwin only, not by native Windows 
programs such as \Code{mprep}).  The temporary files are added to
\Code{tmp} (line 98) and scheduled for deletion at exit with the
\Code{trap} statement in line 105.  The \Code{-mwindows} flag (line
107) adds the Windows system libraries.

A somewhat regrettable feature is that executables produced using Cygwin 
compilers and libraries require Cygwin (or at least \Code{cygwin1.dll}) 
to be installed also on the system the executable is run on.  If one 
does not need \Code{fork}, \Code{wait}, or the \Code{pthread\_*} 
functions (and some few more), it is possible to build executables on 
Cygwin that do not depend on any Cygwin runtime libraries.  There are 
two ways:
\begin{itemize}
\item Either install the \Code{gcc-3} packages (including \Code{g77-3} 
if necessary) and run with 
\begin{verbatim}
CC=gcc-3 CFLAGS=-mno-cygwin
FC=g77-3 FFLAGS=-mno-cygwin
\end{verbatim}
This is restricted to 32-bit, however.

\item Or, install the \Code{mingw} packages for the desired target
(\Code{i686} or \Code{x86\_64}) and work with the targeted versions 
of compilers and binutils, \ie use $h$\Code{-gcc}, $h$\Code{-nm}, 
$h$\Code{-dlltool} instead of \Code{gcc}, \Code{nm}, \Code{dlltool}, 
etc., where $h$ is the `host triplet', \eg \Code{i686-pc-mingw32} or 
\Code{x86\_64-w64-mingw32}.
\end{itemize}

\section{Strings in MathLink}

MathLink has an impressive number of string-related functions which 
differ mainly in how non-ASCII characters are treated.  We will 
concentrate here on two methods only, character strings and byte 
strings.

If the string exchange is known to be in pure ASCII, both methods are
pretty much equivalent in functionality and one can select the more 
convenient one.

Strings from Mathematica should always be considered immutable 
(\Code{const}).  If it becomes necessary to modify such a string, make a 
copy before.  A string read with one of the \Code{MLGet*String} 
functions needs to be de-allocated after use with 
\Code{MLRelease*String}.

\subsection{Character strings}

Character strings are ordinary null-terminated 7-bit C strings.  
Non-ASCII characters (not just accented characters but greek letters, 
mathematical symbols, special punctuation marks, etc.) are encoded by 
Mathematica as escape sequences, such as \verb=\[Alpha]= (`$\alpha$', 8 
characters).

Character strings are the method of choice if one receives a string from 
Mathematica, a function or symbol name, say, and uses that string only 
in the communication with Mathematica in a transparent way, \ie without 
`looking into' it.

For example, the MathLink function template might include a user-defined 
inspector function for debugging (\Code{Identity} if none) to be wrapped 
around (a part of) the result.  The name of that function would have no 
meaning to the MathLink program other than that it gets sent back around 
the right expression.  In this case the MathLink programmer would be 
happy to let Mathematica encode the string in whatever way it needs to 
recognize it as the same later.

The MathLink API functions for character strings are \Code{MLGetString} 
and \Code{MLPutString}.  In the template definition use \Code{String} or 
\Code{Symbol} (depending on the desired pattern matching) and in the 
function declaration \Code{const char *}.

\subsection{Byte strings}

Byte strings are 8-bit character arrays plus a length, \ie are not 
terminated with a special character.  This is conceptually very similar 
to a Fortran string (see App.~\ref{app:fstrings}).  In C one would 
operate on them with the \Code{mem*} family of functions (\Code{memcpy}, 
\Code{memchr}, etc.).

The advantage of byte strings is that the characters map 1:1 onto C or 
Fortran strings (no escape sequences, no variable-length characters as
in UTF-8).  On the other hand, a single byte is not wide enough to hold 
an arbitrary Mathematica character and thus the \Code{MLGetByteString}
function has one argument specifying an 8-bit substitute for characters 
wider than 8 bits.

The MathLink API functions for byte strings are \Code{MLGetByteString} 
and \Code{MLPutByteString}.  In the template definition use 
\Code{ByteString}, in the function declaration \Code{const unsigned char 
*, const int}.

\section{\Code{stdout} and \Code{stderr}}

The typical physicist's practice of writing error, warning, progress 
messages etc.\ on \Code{stdout} (file descriptor 1, Fortran unit 6) or 
\Code{stderr} (file descriptor 2, Fortran unit 0) is not very effective 
in a MathLink program, for \Code{stdout} is suppressed entirely and 
\Code{stderr} appears on the terminal only if running the Mathematica 
Kernel directly (no Frontend).  Silently dropping messages can be 
anything from not helpful to outright dangerous.

An easy though somewhat clumsy solution is to redirect \Code{stdout} 
and/or \Code{stderr} to a file instead.  More elegant is to capture the 
output and send it to Mathematica for display.  Even more flexible is to 
let the user specify a file into which to write the output and recognize
\eg \Code{"stdout"} as a special name in the case of which the output is
sent to Mathematica.

Capturing the output requires a second thread which reads the
\Code{stdout} output generated by the main thread through a pipe and 
sends it to the Kernel.  First we need a few global variables and a copy 
of the original \Code{stdout} file descriptor 1:
\begin{listing}{1}
static int stdoutorig;
static int stdoutpipe[2];
static pthread_t stdouttid;
static int stdoutthr;

static inline void IniRedirect() {
  int fd;
  do fd = open("/dev/null", O_WRONLY);
  while( fd <= 2 );
  close(fd);
  stdoutorig = dup(1);
}

int main(int argc, char **argv) {
  IniRedirect();
  return MLMain(argc, argv);
}
\end{listing}
So as not to overlap with standard file descriptors in the I/O 
redirection later, we open \Code{/dev/null} as many times as it takes to 
obtain a file descriptor not in 0, 1, 2 (lines 8--10), in case 
Mathematica closed any of them.

Capturing is set off by invoking \Code{BeginRedirect} at the beginning 
of the MathLink function in which the messages are generated:
\begin{listingcont}
static inline void BeginRedirect() {
  stdoutthr = pipe(stdoutpipe) != -1 &&
    pthread_create(&stdouttid, NULL,
      MLstdout, NULL) == 0;
  if( !stdoutthr ) stdoutpipe[1] = 2;
  dup2(stdoutpipe[1], 1);
  close(stdoutpipe[1]);
}
\end{listingcont}
In the (somewhat hypothetical) case that \Code{pipe} or 
\Code{pthread\_create} fail, we fall back on the original \Code{stderr} 
(2) so that at least on a terminal there is a chance of seeing the 
output (line 22).  Alternately one could exit with an error code here.  
Then we connect \Code{stdout} (1) to the write end of the new pipe 
(lines 23--24).

The thread function \Code{MLstdout} collects all output in a buffer 
and sends it to Mathematica for display only at the end.  Depending on 
the typical running time and message volume of the underlying function 
one could also send output back line by line, when a certain buffer 
volume is reached, or similar.
\begin{listingcont}
static void *MLstdout(void *dummy) {
  static unsigned char *buf = NULL;
  static long size = 0;
  enum { unit = 10240 };
  long len = 0, n = 0;

  do {
    len += n;
    if( size - len < 128 )
      buf = realloc(buf, size += unit);
    n = read(stdoutpipe[0],
      buf + len, size - len);
  } while( n > 0 );

  if( len ) {
    MLPutFunction(stdlink,
      "EvaluatePacket", 1);
    MLPutFunction(stdlink,
      "WriteString", 2);
    MLPutString(stdlink, "stdout");
    MLPutByteString(stdlink, buf, len);
    MLEndPacket(stdlink);
    MLNextPacket(stdlink);
    MLNewPacket(stdlink);
  }
  return NULL;
}
\end{listingcont}
The output buffer \Code{buf} stays allocated, it usually needs just a 
few kilobytes.  If space runs low, \Code{buf} grows in units of 10 
kbytes (lines 34--35).

Due to the static variables, \Code{MLstdout} is not reentrant but this
is not really necessary, either: to redirect \eg \Code{stderr}, too, 
just connect descriptor 2 to \Code{stdoutpipe[1]} as well, similar to
lines 23--24, rather than create another thread.

The \Code{EndRedirect} function finally cancels the redirection (line 
55) which automatically closes the pipe and causes the reader thread to 
wrap up (lines 40--51) such that it can be joined (line 57).
\begin{listingcont}
static inline void EndRedirect() {
  void *ret;
  dup2(stdoutorig, 1);
  if( stdoutthr )
    pthread_join(stdouttid, &ret);
}
\end{listingcont}
Mathematica is still waiting for the function results by the time 
\Code{EndRedirect} is called, which is why the \Code{WriteString} (line 
43) has to be issued out-of-band in an \Code{EvaluatePacket} (line 41).  
The final \Code{MLNextPacket} and \Code{MLNewPacket} (lines 48--49) 
discard the \Code{WriteString} return value (\Code{Null}).

Note that the transfer of redirected output to Mathematica happens in a 
well-defined sequence, \ie there is no race condition here: it is 
triggered by the \Code{dup2} in line 55 and guaranteed to be finished by 
the time \Code{pthread\_join} returns.  The function's return value is 
transmitted strictly after the \Code{EndRedirect}, so there is no way 
the communication with Mathematica could be upset.

If such a sequence cannot be taken for granted, \eg if \Code{MLstdout} 
is rearranged to send output after each \Code{read} while the main 
thread engages in more communication with the Mathematica Kernel in the 
meantime, it must be enforced using a mutex around the transmission of 
the EvaluatePackets (\eg around lines 41--49 and corresponding ones in 
the main thread).

A final word on buffering: Fortran maintains its own I/O buffers over 
which the C program has no control, thus the message output may not be 
completely transferred by the time the Fortran routine returns. The 
simplest and most portable solution is to explicitly flush unit 6 before 
cancelling the redirection.  Fortran's \Code{flush} subroutine is 
considered an intrinsic by some compilers and therefore cannot portably 
be called from C directly, so a trivial wrapper is necessary:
\begin{verbatim}
   subroutine fortranflush()
   call flush(6)
   end
\end{verbatim}
Add \Code{fortranflush\_();} to \Code{EndRedirect} before the
\Code{dup2} statement and \Code{fortranflush.o} to the \Code{mcc} 
command line.  Alternately, the call to \Code{flush} may be
added to each Fortran routine directly.

For pure C functions a simple \Code{fflush(stdout)} suffices, of course.

\section{Summary}

The preceding sections have collected the information needed to get 
MathLink to work portably across at least the more popular platforms 
currently available, Linux, Mac OS, and Windows.

It is not easy to avoid the impression that this requires more 
workarounds than actual code.  The quality of the MathLink SDK has 
improved somewhat over the versions but is still far from perfect.  
On Mac OS and Windows it would in fact seem that MathLink virtually
eschews any cooperation with the user, which may be connected to the 
fact that these are commercial platforms where users typically do 
not (`are not meant to') build their own executables.

On the positive side it has to be pointed out that providing Mathematica 
connectivity to a piece of C or Fortran code opens up fantastic new 
possibilities for interactive use (think of functions like 
\Code{Manipulate}) and combination with Mathematica's sophisticated 
functions (if unconvinced, try doing a \Code{ContourPlot} in Fortran, 
for example).  For users not sufficiently familiar with C or Fortran, it 
makes these functions available at all.

The only significant work is writing interfacing code in a \Code{.tm} 
program.  Apart from that, a package author only needs to add the 
\Code{fcc} and \Code{mcc} substitute scripts and tweak the makefile as 
described in Sect.~\ref{sect:mcc}.  All things considered, this is a 
fairly moderate effort.

The shell scripts together with demo code are available from 
\Code{http://feynarts.de/mathlink/}.  They can be witnessed `in action' 
in the packages LoopTools \cite{lt}, Cuba \cite{cuba}, FeynHiggs 
\cite{fh}, Diag \cite{diag}, and in FormCalc's Mathematica interface
\cite{fcmma}.

Comments, improvements, and bug-fixes are welcome at 
\Code{hahn@feynarts.de}.

\begin{appendix}

\section{C--Fortran interfacing}

MathLink's native tongue is C so in order to link Mathematica with 
Fortran code one needs to know at least the basics of C--Fortran 
interfacing.  Fortran is less flexible in its calling conventions, thus 
in general the C program has to adapt, not the Fortran program.

\subsection{Function names}

Fortran names (subroutines, functions, common blocks, block data) are 
lowercased and an underscore is appended by the time they end up in the 
object file.  The very few compilers which do not add an underscore 
(HP-UX's fort77, for example) should largely be extinct by now.  Steer 
clear of underscores in Fortran names, as compilers have different ways 
of mangling those, \eg some compilers add \emph{two} underscores if the 
Fortran name already contains one (cf.\ gfortran's 
\Code{-fsecond-underscore} option).

Modern C compilers require prototypes for all external routines.  
In C++, the prototype must be wrapped in \verb=extern "C" {...}= to 
suppress C++ name mangling.

Example: \Code{subroutine FOO} in Fortran becomes \Code{void foo\_()}
in C.

In case of problems, check the spelling of the symbols as visible to the 
linker by using \Code{nm} on the Fortran object file (and, for 
comparison, possibly the C object file as well).

\subsection{Function arguments}

Fortran always calls by reference, \ie passes a pointer, not the
variable itself (exceptions only when using the much-deprecated 
\Code{\%VAL}).

Example:
\begin{verbatim}
   subroutine foo(i)
   integer i
\end{verbatim}
becomes \Code{void foo\_(int *i)} in C -- mind the \Code{*}.

Hint 1: Since Fortran passes by reference, and rather indiscriminately 
so (for example, many compilers silently ignore mismatches between 
formal and actual arguments even within the same source file), it is an
extremely good idea to set up prototypes as strictly as possible,
including \Code{const} in all places where the argument should not
be modified (even though Fortran has no way of controlling this).

Hint 2: For portability between C and C++, use the \Code{\_\_cplusplus} 
preprocessor variable around the \Code{extern "C"} bits:
\begin{verbatim}
#ifdef __cplusplus
extern "C" {
#endif

void a0_(double *res, const double *m);
(... possibly more prototypes ...)

#ifdef __cplusplus
}
#endif
\end{verbatim}

Hint 3: It makes life a lot easier to wrap Fortran functions in C inline 
functions, with proper C calling conventions.  The wrapper function can 
be used just as a regular C function and `inline' means there is no 
extra calling overhead, \eg
\begin{verbatim}
static inline double A0(const double m) {
  double res;
  a0_(&res, &m);
  return res;
}
\end{verbatim}
Take care that the \Code{static} attribute pertains to the wrapper 
function only, as the Fortran function most certainly does not have file 
scope but comes from an external library.

\subsection{Return values}

Prefer subroutines over functions in Fortran, especially if the return 
value is not an \Code{integer} or \Code{double precision} 
(\Code{real*8}).  Conventions vary most notably for \Code{double 
complex} functions.  If a function is required on the Fortran side \eg 
because of an API, add a subroutine wrapper:
\begin{verbatim}
   double complex foo(args...)
   ...
   end

   subroutine foowrapper(res, args...)
   double complex res, foo
   external foo
   res = foo(args...)
   end
\end{verbatim}

\subsection{Data types}
\label{app:ftypes}

Most scalar types have an obvious counterpart in C, \eg
\begin{itemize}
\item\Code{integer} (\Code{integer*4}) --- \Code{int}, \\
  \Code{integer*2} --- \Code{short}, \\
  \Code{integer*8} --- \Code{long long int},
\item\Code{double precision} (\Code{real*8}) --- \Code{double}, \\
  \Code{real} (\Code{real*4}) --- \Code{float},
\item\Code{double complex} (\Code{complex*16}) \\
  --- \Code{double complex} (C99), \\
  --- \verb=struct { double re, im; }= (C89), \\
  --- \Code{std::complex<double>} (C++),
\item\Code{character} --- \Code{char} \\
  (but for strings see below).
\end{itemize}
There is no portable equivalent of \Code{logical} in C, however; it is 
better to use an integer in Fortran instead.  If you must use
\Code{logical}, interface with an \Code{int} and test for the lowest bit 
only.

These correspondences can also be coded with \Code{typedef} statements.  
Not only does this make the Fortran types stand out visually, but the
compiler will automatically add casts or warn about incompatibilities 
when mixing with other C types.  Note the \Code{const} versions for 
strict prototyping.
\begin{listing}{1}
typedef int INTEGER;
typedef const INTEGER CINTEGER;
typedef double REAL;
typedef const REAL CREAL;
typedef struct { REAL re, im; } COMPLEX;
typedef const COMPLEX CCOMPLEX;
typedef char CHARACTER;
typedef const CHARACTER CCHARACTER;
\end{listing}
For portability between C and C++, for example in header files, one may 
want to provide wrapper code for the homogeneous treatment of complex 
numbers:
\begin{listingcont}
#define Real double 
#define ToReal(r) (r)

#ifdef __cplusplus

#include <complex>
typedef std::complex<Real> Complex;
#define ToComplex(c) \
  Complex(ToReal((c).re), ToReal((c).im))
#define ToComplex2(r,i) \
  Complex(r, i)
#define Re(x) std::real(x)
#define Im(x) std::imag(x) 

#else

#include <complex.h>
typedef Real complex Complex;
#define ToComplex(c) \
  (ToReal((c).re) + I*ToReal((c).im))
#define ToComplex2(r,i) \
  (r + I*(i))
#define Re(x) creal(x)
#define Im(x) cimag(x)

#endif
\end{listingcont}
Referring to reals indirectly through the \Code{Real} and \Code{ToReal} 
macros is useful for switching to a different precision (\Code{float}, 
\Code{long double}; see App.~\ref{app:quad}).

\subsection{Common blocks}

Common blocks map onto C structs, with the members in the same order.  
The \Code{struct} should be declared \Code{extern} to prevent 
instantiation in C.  In C++, \Code{extern "C"} must be used in addition 
to the \Code{extern} (storage class modifier) because of name mangling,
as with the function prototypes.

Padding might be an issue if common members are ordered unsuitably for 
alignment, \eg
\begin{verbatim}
   integer i
   double precision r
   common /c/ i, r
\end{verbatim}
is not correctly aligned for a 64-bit architecture because the integer 
is a 32-bit quantity.  Ideally, the common block should be reordered 
(widest types first, \eg \Code{double complex} before \Code{double 
precision} before \Code{integer} before \Code{character}).  If this 
is not possible due to API requirements or similar, wrap the common 
definition in
\begin{verbatim}
#pragma pack(push, 1)
extern struct {
  int i;
  double r;
} c_;
#pragma pack(pop)
\end{verbatim}
This switches off padding (in gcc at least) but carries a performance 
penalty, and on RISC platforms such as the Alpha likely triggers
unaligned-access exceptions (in both Fortran and C).

\subsection{Strings}
\label{app:fstrings}

There are no strings in Fortran, only character arrays (padded with 
spaces as necessary), and these are handled specially by the compiler,
\ie differently from other arrays.

The C function gets called with two arguments for every Fortran string: 
a \Code{char *} pointer to the characters, in the same place as the 
string argument in Fortran, and a \Code{const int} following the Fortran 
argument list.  Example:
\begin{verbatim}
   subroutine strfoo(s1, i1, s2, i2)
   character*(*) s1, s2
   integer i1, i2
\end{verbatim}
in C becomes
\begin{verbatim}
void strfoo_(char *s1, int *i1,
  char *s2, int *i2,
  const int s1_len, const int s2_len)
\end{verbatim}
Fortran strings are \emph{not} null-terminated and it is in general, but 
in particular for functions invoked with string literals as arguments, 
not advisable to write a zero-byte into the Fortran string \textit{in 
situ} and hope it won't disturb Fortran later.

The two `clean' options are: use the \Code{mem*} family of functions 
(\Code{memcpy}, \Code{memchr}, etc.)\ which take a length argument and 
do not require a terminating zero-byte, or null-extend the string in 
allocated memory.  The latter is particularly simple in C99/C++ where 
space can easily be allocated on the stack (and is automatically 
de-allocated when the object goes out of scope), \eg
\begin{verbatim}
void sf_(char *s, const int s_len) {
  char sn[s_len + 1];
  memcpy(sn, s, s_len);
  sn[s_len] = 0;
  ...
}
\end{verbatim}
Note that, while the string \Code{sn} is correctly null-terminated now 
and may be worked on with the \Code{str*} functions, it may well include 
trailing spaces as Fortran indicates only the allocated size in 
\Code{s\_len}, not the actual length (sans trailing spaces) of the 
character array.

In the opposite direction, invoking a Fortran function with a string
argument is straightforward, \eg
\begin{verbatim}
fortfoo_(s, strlen(s));
\end{verbatim}
When returning a string of a given length, \ie (over)writing one of the 
subroutine's string arguments, the unused characters should be filled 
with spaces, as in:
\begin{verbatim}
void cfoo_(char *s, const int s_len) {
  ...
  n = strlen(result);
  if( n >= s_len )
    memcpy(s, result, s_len);
  else {
    memcpy(s, result, n);
    memset(s + n, ' ', s_len - n);
  }
}
\end{verbatim}

\section{Required flags and libraries}
\label{app:fldflags}

The rules above are sufficient to obtain object-level compatibility, \ie 
making C and Fortran routines talk to each other.  For successful 
linking one has to add the Fortran compiler's run-time libraries (\eg 
I/O, extended math) to the C command line as well.

To determine the necessary flags, the Fortran compiler is run in verbose 
mode on a test program and all flags relevant to linking are collected 
from the output.  Here is how:
\begin{listing}{1}
getldflags() {
  ldflags="$LDFLAGS"
  while read line ; do
    set -- `echo $line | tr ':,()' '    '`
    case $1 in
    */collect2* | */ld* | ld*) ;;
    *) continue ;;
    esac
    while test $# -gt 1 ; do
      shift   
      case $1 in
      *.o | -lc | -lgcc*)
        ;;
      -l* | -L* | *.a)
        ldflags="$ldflags $1" ;;
      -Bstatic | -Bdynamic | *.ld)
        ldflags="$ldflags -Wl,$1" ;;
      /*)
        ldflags="$ldflags -L$1" ;;
      -rpath*)
        ldflags="$ldflags -Wl,$1,$2"
        shift ;;
      -dynamic-linker)
        shift ;;
      esac
    done
  done
  echo $ldflags
}
\end{listing}
Lines 5--8 select only the linker command lines (note that gcc uses 
\Code{collect2}, not \Code{ld}).  The \Code{while} loop starting in line 
3 picks out the relevant items from the ld command line.

The above shell function is used as in
\begin{verbatim}
LDFLAGS=`f77 -v -o test test.f 2>&1 | \
  getldflags`
\end{verbatim}
where \Code{test.f} is a simple test program, preferably one producing
output so that the I/O libraries are referenced, \eg
\begin{verbatim}
   program hw
   print *, "Hello World"
   end
\end{verbatim}

\section{Extended precision}
\label{app:quad}

MathLink provides a data type by the name of \Code{Real128}.  Contrary 
to expectations, this does not represent a genuine quadruple-precision 
128-bit floating-point number but `only' C's \Code{long double} data 
type which on Intel x86 hardware is usually the 80-bit 
`extended-precision' float, even though C stores this in 12 (in 32-bit) 
or 16 bytes (in 64-bit mode) for alignment.

Those few Fortran compilers that offer higher than double precision 
typically have the \Code{real*16} data type which implements true 
128-bit precision, albeit in software emulation (\ie slow).

The two formats are actually quite similar, the main difference being 
that the most-significant bit of the (normalized) fraction is implicit 
in \Code{real*16} and explicit in \Code{long double} (which can thus 
also represent denormalized fractions):

\noindent
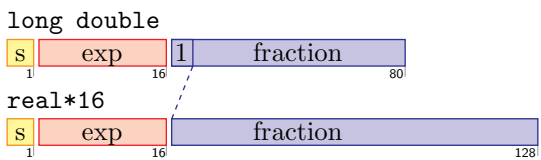
\begin{picture}(200,70)
\SetColor{Gray}
\SetWidth{.2}
\SetOffset(0,10)
\Line(10,0)(10,-4)
\Line(60,0)(60,-4)
\Line(200,0)(200,-4)
\SetOffset(0,40)
\Line(10,0)(10,-4)
\Line(60,0)(60,-4)
\Line(150,0)(150,-4)
\SetColor{Blue}
\SetWidth{.5}
\DashLine(62,-20)(70,0){2}
\SetOffset(0,40)
\Text(0,12)[bl]{\Code{long double}}
\CBox(0,0)(10,10){BurntOrange}{PastelYellow}
\Text(5,0)[b]{s\vphantom{p}}
\Text(10,-1)[tr]{\tiny\textsf{1}}
\CBox(12,0)(60,10){Red}{PastelRed}
\Text(36,0)[b]{exp}
\Text(60,-1)[tr]{\tiny\textsf{16}}
\CBox(62,0)(150,10){Blue}{PastelBlue}
\CBox(62,0)(70,10){Blue}{PastelBlue}
\Text(66,0)[b]{1\vphantom{p}}
\Text(110,0)[b]{fraction\vphantom{p}}
\Text(150,-1)[tr]{\tiny\textsf{80}}
\SetOffset(0,10)
\Text(0,12)[bl]{\Code{real*16\vphantom{p}}}
\CBox(0,0)(10,10){BurntOrange}{PastelYellow}
\Text(5,0)[b]{s\vphantom{p}}
\Text(10,-1)[tr]{\tiny\textsf{1}}
\CBox(12,0)(60,10){Red}{PastelRed}
\Text(36,0)[b]{exp}
\Text(60,-1)[tr]{\tiny\textsf{16}}
\CBox(62,0)(200,10){Blue}{PastelBlue}
\Text(110,0)[b]{fraction\vphantom{p}}
\Text(200,-1)[tr]{\tiny\textsf{128}}
\end{picture}

Conversion can be achieved \eg as follows (note that this code is 
specific to Intel x86 hardware):
\begin{listing}{1}
#pragma pack(push, 1)
typedef union {
  long double r10;
  struct {
    unsigned long long frac;
    unsigned short exp;
  } i10;
  struct {
    char zero[6];
    unsigned long long frac;
    unsigned short exp;
  } i16;
  unsigned long long i8[2];
} real16;
#pragma pack(pop)

static inline real16 ToREAL(long double r) {
  real16 new;
  new.i8[0] = 0;
  new.i16.frac =
    ((real16 *)&r)->i10.frac << 1;
  new.i16.exp = ((real16 *)&r)->i10.exp;
  return new;
}

static inline long double ToReal(real16 r) {
  real16 new;
  const long long z = r.i16.frac |
    (r.i16.exp & 0x7fff);
  new.i10.frac = (r.i16.frac >> 1) |
    ((z | -z) & 0x8000000000000000);
  new.i10.exp = r.i16.exp;
  return new.r10;
}

#define Real long double
typedef real16 REAL;
\end{listing}
One needs to apply \Code{ToREAL} when transferring arguments from C to 
Fortran, and \Code{ToReal} when transferring from Fortran to C.  
Defining \Code{Real} as in line 36 automatically upgrades the 
\Code{Complex} data type declared in App.~\ref{app:ftypes}, too.

Double-precision Fortran code can be upgraded to quadruple precision 
almost automatically if a few rules are observed, through compiler 
flags like \Code{-r16} (ifort).  This can be very helpful to find out 
whether a wrong result comes from a programming mistake (bad algebra)
or from \eg accumulation of round-off errors (bad numerics).  The rules:
\begin{itemize}
\item Use \Code{double precision}/\Code{double complex}, not
  \Code{real*8}/\Code{complex*16}, the latter are not affected
  by the automatic \Code{-r16} promotion.

\item Use generic functions only, \eg use \Code{abs}, not \Code{dabs}.
  To force a type conversion, make an explicit cast, \eg use 
  \Code{sqrt(DCMPLX(x))}, not \Code{cdsqrt(x)} (on \Code{DCMPLX} 
  see next item).

\item Despite their name, \Code{real}, \Code{imag}, \Code{conjg}, and 
  \Code{cmplx} are not generic but single-precision real 
  (\Code{real*4}) functions and thus not promoted automatically,
  either.

  Solution: use the explicitly double-precision functions \Code{DBLE},
  \Code{DIMAG}, \Code{DCONJG}, and \Code{DCMPLX} in the original code.
  Choose identical capitalization (\eg all-caps as given
  here so as to stand out) and have the preprocessor substitute
  the double- by quadruple-precision variants by adding the following 
  flags to the command line:
\begin{verbatim}
-DDBLE=QEXT -DDCONJG=QCONJG \
-DDCONJG=QCONJG -DDCMPLX=QCMPLX
\end{verbatim}
  Note: The file extension should be \Code{.F} in this case, not 
  \Code{.f}, otherwise the preprocessor will not be invoked 
  automatically.
\end{itemize}

\end{appendix}

\raggedright


\begin{thebibliography}{99}

\bibitem{lt}
T.~Hahn, M.~Perez-Victoria, Comput.\ Phys.\ Commun.\ \textbf{118}
(1999) 153--165 [hep-ph/9807565].

\bibitem{cuba}
T.~Hahn, Comput.\ Phys.\ Commun.\ \textbf{168} (2005) 78--95
[hep-ph/0404043].

\bibitem{fh}
M.~Frank, T.~Hahn, S.~Heinemeyer, W.~Hollik, H.~Rzehak, G.~Weiglein,
JHEP \textbf{0702} (2007) 047 [hep-ph/0611326].

\bibitem{diag}
T.~Hahn, physics/0607103.

\bibitem{fcmma}
T.~Hahn, Comput.\ Phys.\ Commun.\ \textbf{178} (2008) 217
[hep-ph/0611273].

\end{thebibliography}
\end{document}